\documentclass[sigconf]{acmart}

\AtBeginDocument{%
  \providecommand\BibTeX{{%
    \normalfont B\kern-0.5em{\scshape i\kern-0.25em b}\kern-0.8em\TeX}}}

\copyrightyear{2023}
\acmYear{2023}
\setcopyright{acmlicensed}
\acmConference[CHI '23]{Proceedings of the 2023 CHI Conference on Human Factors in Computing Systems}{April 23--28, 2023}{Hamburg, Germany}
\acmBooktitle{Proceedings of the 2023 CHI Conference on Human Factors in Computing Systems (CHI '23), April 23--28, 2023, Hamburg, Germany}
\acmPrice{15.00}
\acmDOI{10.1145/3544548.3581239}
\acmISBN{978-1-4503-9421-5/23/04}

\usepackage{tabularx}
\usepackage{booktabs}
\usepackage{graphicx}
\usepackage{hyperref}
\usepackage{hhline}

\usepackage{subcaption}
\usepackage{enumitem}

\usepackage{float} 
\usepackage{soul,color}
\usepackage[utf8]{inputenc}

\usepackage{longtable}
\usepackage{chngcntr}
\usepackage{xcolor}
\definecolor{editCol}{rgb}{0.0, 0.0, 0.0}
\newcommand{\edit}[1]{{\textcolor{editCol}{#1}}}
\usepackage{pbox}
\usepackage{multirow}
\usepackage{array}
\usepackage{makecell}
\usepackage{afterpage} 
\usepackage[title]{appendix}
\usepackage{subcaption,graphicx}
\usepackage{ragged2e}

\newcolumntype{L}[1]{>{\raggedright\let\newline\\\arraybackslash\hspace{0pt}}m{#1}}
\newcolumntype{C}[1]{>{\centering\let\newline\\\arraybackslash\hspace{0pt}}m{#1}}
\newcolumntype{R}[1]{>{\raggedleft\let\newline\\\arraybackslash\hspace{0pt}}m{#1}}

\begin{document}

\title[Co-monitoring In-home Emergencies]{Co-designing Community-based Sharing of Smarthome Devices for the Purpose of Co-monitoring In-home Emergencies }


\author{Leena Alghamdi}
\email{Leenaalghamdi@knights.ucf.edu}
\orcid{0000-0003-2102-9155}
\affiliation{%
  \institution{University of Central Florida}
  \city{Orlando}
  \state{Florida}
  \postcode{32826}
  \country{USA}
}

\author{Mamtaj Akter}
\email{Mamtaj.Akter@vanderbilt.edu}
\orcid{0000-0002-5692-9252}
\affiliation{%
  \institution{Vanderbilt University}
  \city{Nashville}
  \state{Tennessee}
  \postcode{37212}
  \country{USA}
}

\author{Jess Kropczynski}
\email{Jess.Kropczynski@uc.edu}
\orcid{0000-0002-7458-6003}
\affiliation{%
  \institution{University of Cincinnati}
  \streetaddress{P.O. Box 1234}
  \city{Cincinnati}
  \state{OH}
  \postcode{45221}
  \country{USA}
}

\author{Pamela J. Wisniewski}
\email{Pamela.Wisniewski@vanderbilt.edu}
\orcid{0000-0002-6223-1029}
\affiliation{%
  \institution{Vanderbilt University}
  \city{Nashville}
  \state{Tennessee}
  \postcode{37212}
  \country{USA}
}

\author{Heather Lipford}
\email{Heather.Lipford@uncc.edu}
\orcid{0000-0002-5261-0148}
\affiliation{%
  \institution{University of North Carolina, Charlotte}
  \city{Charlotte}
  \state{North Carolina}
  \postcode{28223}
  \country{USA}
 }


\renewcommand{\shortauthors}{Alghamdi et al.}

\begin{abstract}
We conducted 26 co-design interviews with 50 smarthome device owners to understand the perceived benefits, drawbacks, and design considerations for developing a smarthome system that facilitates co-monitoring with emergency contacts who live outside of one's home. Participants felt that such a system would help ensure their personal safety, safeguard from material loss, and give them peace of mind by ensuring quick response and verifying potential threats. However, they also expressed concerns regarding privacy, overburdening others, and other potential threats, such as unauthorized access and security breaches. To alleviate these concerns, participants designed flexible and granular access control and fail-safe back-up features. Our study reveals why peer-based co-monitoring of smarthomes for emergencies may be beneficial but also difficult to implement. Based on the insights gained from our study, we provide recommendations for designing technologies that facilitate such co-monitoring while mitigating its risks.
\end{abstract}


\maketitle

\section{Introduction}
Smarthomes are becoming increasingly common as the number of homes with smart devices in the world is expected to exceed 350 million by 2023 ~\cite{cycles_topic_2022}. According to a recent report from Juniper Research, by 2025, there will be approximately 13.5 billion smarthome devices in active use ~\cite{price_smart_2020}.
Smarthome devices can be utilized to provide a safe and secure environment to smarthome owners by protecting them from various threats, such as fire, water leaks, and theft. Emergency situations disrupt the lives of people and can cause long-term consequences for one's home and personal safety. For instance, according to the National Fire Protection Association, in a five-year period, in-home fires caused an annual average of 2,620 deaths, 11,070 injuries, and \$7.3 billion in property damage \cite{noauthor_home_2021}. As such, people have begun to use smarthome security systems for emergency monitoring. Professional monitoring companies, such as ADT \cite{noauthor_best_2022}, have extended their offerings to include smarthome devices, while smarthome monitoring companies, like Ring \cite{siminoff_history_2014}, have emerged over the course of the last decade. Smarthome security systems provide benefits over traditional in-home security systems, as smarthome owners can become aware of issues in real-time that have been reported by home automation technology, increasing the chances of early mitigation ~\cite{noauthor_what_2018}.

However, these smarthome security systems have several downsides. For instance, alarm systems are expensive and sometimes complicated to install, including equipment, installation, and security service subscriptions ~\cite{majumder_smart_2017}. Moreover, most security systems are prone to false alarms that could be costly to the homeowner if emergency services are summoned unnecessarily. According to the National Fire Protection Association, the United States fire department responds to over 2.2 million false alarms \cite{httppragmaticmatescom_costs_2016} on average each year, with one in twelve of the false alarms triggered by a home security system. False alarms that trigger emergency response can cost authorities over \$2 billion a year \cite{corp_real_2019} of taxpayer dollars. In other cases, when home security systems are not professionally monitored, notifications of real emergencies may be missed by homeowners, resulting in material loss and/or personal harm ~\cite{raja_inclusive_2013}.

Smarthome owners may not want to pay the costly fees associated with professional security monitoring but are not always available to perform the emergency monitoring of their homes themselves. As such, recent research by Tabassum et al. ~\cite{tabassum_smart_2020} uncovered that almost half (47.8\%) of smarthome device owners in their study shared devices with people outside of their homes for various purposes. While many of these use cases did not involve in-home emergency monitoring (e.g., sharing smart doorbells and smart locks for easy access to the house for tasks like pet-sitting), 55\% of their participants said they shared or would share their smarthome devices with trusted individuals outside their homes to increase the safety and security of their homes. While this work uncovered smarthome owners' propensity and desire to share their smarthome devices for these purposes, there is a knowledge gap in how to design systems to support this type of in-home emergency co-monitoring. Therefore, we build upon this recent work by deeply exploring potential benefits, drawbacks, and design considerations of sharing smarthome devices with others for in-home emergency co-monitoring\edit{, with the end goal of this research to arrive at novel insights for developing a co-monitoring mobile app for smarthome device owners.} In this paper, we use the term "co-monitoring" to refer to the concept of how trusted contacts, e.g., friends and family who live outside of the home, help homeowners watch over their houses through smarthome devices, such as security cameras and motion detectors, that are shared with them. We designed our study to answer the following high-level research questions:

\begin{itemize}
\item \textbf{RQ1:} \textit{What are the motivations and benefits of sharing smart-home devices with other people outside the home for in-home emergency co-monitoring purposes?}
\item \textbf{RQ2:} \textit{What are the drawbacks of sharing smarthome devices with other people outside the home for in-home emergency co-monitoring? What are the barriers which would prevent smarthome device owners from sharing?}
\item \textbf{RQ3:} \textit{What are the important design considerations and users' needs for designing a smarthome emergency co-monitoring system?}
\end{itemize}

To answer these research questions, we conducted 26 co-design interviews with 50 smarthome device owners (24 dyads of two participants plus two sessions with individuals). \edit{We used a multi-modal methodological approach, where} we presented scenarios related to smarthome emergency situations and asked questions to prompt participants to reflect on their perceptions and concerns in that context (RQ1 \& RQ2). Participants were then asked to design features for an emergency co-monitoring app for sharing access to their smarthome devices with other people outside their homes (i.e., emergency contacts) for in-home emergency co-monitoring (RQ3). 

Overall, we found that most participants saw benefits in co-monitoring their smarthome devices with trusted emergency contacts, as it would help them ensure personal safety, safeguard from material loss, and provide peace of mind by ensuring quick emergency response and verifying potential threats (RQ1). However, having their emergency contacts watch over their homes would also potentially invade their privacy and overburden emergency contacts. They also felt that co-monitoring with outsiders might open their homes to new threats, such as unauthorized access and security breaches (RQ2). To alleviate these concerns, participants designed co-monitoring features (RQ3) such that they could have flexible and granular controls on the access of emergency contacts, by limiting access duration, devices, and/or device activities. Smarthome device owners also suggested features to improve emergency response by providing unignorable notifications, features to safeguard against improper use, and including automated emergency calls.

Our research makes important contributions to the internet of things and networked privacy research communities by examining whether co-monitoring smarthome devices with trusted emergency contacts could potentially help device owners manage in-home emergencies. Our results represent a first step in this exploration, with this study focusing on low-risk populations and settings. Still, our study encourages the research community to think critically about the design of smarthome co-monitoring tools for device owners and demonstrates how we might leverage such co-monitoring mechanisms to ensure safety and security in the context of in-home emergencies, as well as mitigating for challenges. Therefore, we make the following unique research contributions: 

\begin{itemize}
\item Through presenting hypothetical scenarios of emergencies in smarthomes, we inspired device owners to imagine such emergencies at their homes and brainstorm what they could do to avoid them. 

\item We gained empirical insights into the potential benefits of sharing smarthome devices with emergency contacts to manage emergencies, but also the \edit{serious privacy} concerns that might make sharing smarthome devices with outsiders challenging to implement.

\item We present design-based features from our co-design sessions that would better support smarthome device owners in sharing their devices with trusted emergency contacts, enabling them to respond to unexpected in-home emergencies. 
\end{itemize}


\section{Background}
We place our study within three main streams of research: 1) Benefits and challenges of using smarthome systems for emergency monitoring, 2) Smarthome systems access control, and 3) Sharing smarthome devices beyond the home for emergency monitoring.

\subsection{Benefits and challenges of using smarthome systems for emergency monitoring}

 Over the past decade, researchers have proposed how to build and utilize smarthome systems to detect a range of dangerous situations, such as fires, water leaks, gas leaks, and break-ins (e.g. ~\cite{ali_solution_2022,saeed_iot-based_2018,sarhan_systematic_2020,salhi_early_2019} ). In addition, a large number of commercial sensors are now available to aid in monitoring the safety and security of a home, including smart smoke and carbon monoxide detectors, water and temperature sensors, contact sensors, motion detectors, air quality monitors and sound sensors. According to a PC Mag survey conducted in 2018 ~\cite{Moscaritolo_which_2018}, 17\% of respondents already had one or more home security devices, while 14\% planned to purchase one in the near future. Respondants cited peace of mind, increased security, and remote home monitoring as benefits of such devices. As such, commercial and professionally monitored home security systems, such as ADT \footnote{https://www.adt.com} and Simplisafe \footnote{https://www.simplisafe.com}, are now incorporating smarthome devices for detecting and mitigating a range of in-home emergency situations. Recognizing the benefits, some insurance companies may offer discounts for having a professionally monitored home security system ~\cite{noauthor_will_2022}.


However, smarthome security and safety systems have their drawbacks. Professionally installed and/or monitored systems can be expensive, requiring large up-front installation costs and monthly service fees. Even with professional monitoring, false positives can occur, wasting the time and resources of emergency responders ~\cite{httppragmaticmatescom_costs_2016, corp_real_2019} and worrying or annoying homeowners. As a result, many homeowners are purchasing and utilizing their own devices, choosing the subset of devices they desire, so they can monitor the safety and security of their homes themselves. This allows for more flexible configuration, portability, and cost savings ~\cite{noauthor_pros_2022, 10.1007/978-3-031-05563-8_24}. However, homeowners then become responsible for their own emergency response. They must successfully install and configure different devices, often from different manufacturers using different app interfaces, without professional assistance. Further, they must then notice, understand, and respond appropriately to emergency alerts in a timely manner ~\cite{noauthor_pros_2022}. Thus, smarthome device owners may appreciate sharing this task with others within or outside of the home. \edit{Yet, sharing one's smarthome devices with people outside of one's home may also create new challenges related to privacy and security. For instance, Chouhan et al.'s framework ~\cite{chouhan_co-designing_2019, akter_CO-oPS_2022} of community oversight for privacy and security highlighted trust as a key component in co-monitoring relationships. Researchers within the HCI research community have also surfaced the concern that surveillance-based smarthome systems can be used to facilitate harm to vulnerable users, such as victims of intimate partner violence (c.f., ~\cite{leitao_anticipating_2019, parkin_usability_2020, rodriguez-rodriguez_towards_2020, slupska_threat_2021, mcdonald_privacy_2022, noauthor_ipv_2022}). Therefore, assessing the potential privacy and security threats of smarthome co-monitoring, not only the potential benefits, was a key goal of our study. One approach for mitigating privacy and security concerns is through the careful design of smarthome device access controls, which we unpack further in the next section.}

\subsection{Smarthome systems access controls}
Smarthome devices are frequently utilized by people beyond the individual who purchased and installed the device ~\cite{sikder_whos_2022}, including others living within a home, e.g. spouses, roommates ~\cite{lekakis_dont_2012}, and children ~\cite{schechter_user_2013}), guests in the home ~\cite{johnson_usability_2009}, as well as family and friends outside of the home~\cite{tabassum_smart_2020}. Thus, many smart devices provide some capability to share devices and data access with secondary users, people other than the user who initially installed the device. One reason device owners might share is to enable other home residents or community members to assist in monitoring the various notifications coming from smarthome security devices. Thus, we first highlight prior research in access control for smarthome devices more generally before turning to the specific context of emergency response.

Research has demonstrated that users can have complex access control needs within the smarthome. For example, through a large-scale vignette study, He et al. identified that participants' desired access policies differed significantly based on the relationship with the secondary users, contextual factors such as time and location, as well as device type and even specific device capabilities ~\cite{he_rethinking_2018}. Additional studies have also identified that access control policies may depend on time, location, user roles, and device and data types ~\cite{sikder_kratos_2020,zeng_understanding_2019}. Many smarthome devices, as well as device integration platforms (e.g. Samsung SmartThings, Apple Home Hub), provide features for sharing devices with multiple people. However, some provide the same level of access to the secondary users as the device owner ~\cite{sikder_whos_2022}. This allows secondary users to not only have full control over a smart device, but also to reconfigure devices such as adding or deleting users (even the homeowner themselves). Many devices offer coarse-grained control. For example, the Ring doorbell offers a feature that allows the owner to add a user to the doorbell, providing access to a predefined set of capabilities ~\cite{noauthor_controlling_2022}. However, a challenge with these features is that users have difficulty determining exactly what is shared ~\cite{tabassum_i_2019}. Few existing devices and smarthome integration platforms offer any sort of fine-grained access control, such as time- or location-based configuration ~\cite{mare_consumer_2019}. Whether secondary users are truly given full access or not, users may need assume they are. As a result, many only share with their most trusted friends or family members ~\cite{tabassum_i_2019, akter-from-2022}. Some simply bypass these basic access control systems altogether and just share full account credentials to share access ~\cite{tabassum_i_2019}. 

Several projects have proposed more complex access control systems for smarthome devices. For example, Sikder et al. introduced KRATOS+ as a flexible multi-user access control system for multiple devices and demonstrated it can express and resolve access conflicts in real-world scenarios ~\cite{sikder_whos_2022}. Zeng et al. ~\cite{zeng_understanding_2019} developed a prototype smarthome app that provided fine-grained access control that included support for role-based, location-based, and reactive access controls. However, in a month-long field study, they found little use of nuanced access control, either because of the complexity of setting up the policy or the strong trust among the household members. Their results demonstrate that as control systems become more complex, that complexity may result in users feeling less in control ~\cite{randall_living_2003}.

\subsection{Sharing smarthome devices beyond the home for emergency monitoring}
Much of the above research has examined sharing devices with users within the home -- other residents or visitors. However, Tabassum et el. ~\cite{tabassum_smart_2020} found that nearly half of their survey participants were already sharing their smart devices with others outside of their homes. Brush et al. first introduced the notion of a digital neighborhood watch, where users would share their outdoor security cameras with neighbors for shared monitoring of a community ~\cite{brush_digital_2013}. They found that willingness to share with individual neighbors was based on a trusted relationship rather than proximity, and users were generally unwilling to share direct camera access with others. However, Tabassum et al. is the only recent study we are aware of examining the motivations and needs of sharing modern smart devices with those outside of the home ~\cite{tabassum_smart_2020}. Participants reported sharing to enable a trusted set of family and friends to help manage the safety and security of the home and its occupants. Use cases included checking on pets and houses while owners were away, allowing remote access to the home, communicating with home occupants, and finally, monitoring and responding to emergencies. The study identified that current systems do not adequately support such sharing needs. Systems need to provide finer-grained access control so users would be comfortable sharing with less trusted contacts, such as neighbors, as well as additional time- and event-based access control to better support the use cases raised by participants ~\cite{tabassum_smart_2020}. \edit{Thus, prior work has established that people are interested in sharing smart home devices for emergency monitoring; yet, relatively little work has been done to study how we can actually enable such interactions in real-world smarthome systems, while adequately addressing privacy concerns that would arise. Our study fills this gap.}

\section{Methods}
\edit{A co-design interview study was well-suited for answering our research questions, as it combines a retrospective approach to elicite users' thoughts and experiences of current usage practices, with a more generative methodology (i.e., co-design), where researchers work directly with potential users of a system to co-create features that would fulfill users' anticipated needs ~\cite{muller_participatory_1993, spinuzzi_methodology_2005}.} \edit{Muller and Druin ~\cite{muller_participatory_1993} described co-design as a space where users can challenge technology developers' underlying assumptions, generate new ideas, and co-create better design solutions than developers would have envisioned on their own. Co-design gives researchers a unique opportunity to interrogate their own ideas (in this case, community-based co-monitoring of smarthomes for emergencies) and, more importantly, to take a generative approach to extend beyond our preconceptions ~\cite{muller_participatory_1993}.} \edit{Moreover, we decided to conduct the co-design interview with pairs of participants, known as a paired depth interview, to reflect privacy as a dialectical interpersonal process ~\cite{altman_dialectic_1981}. The paired-depth interview enabled us to delve thoroughly into each person's experience while using any similarities and differences to further investigate the topic. Both interviewees could also equally participate in the discussion that occured, and the back-and-forth discussion provided more complete data ~\cite{cartwright_using_2016}.}
\edit{Below, we provide the details of our recruitment strategy, the study methods, the data analysis approach, and the demographic profiles of our participants.}

\subsection{Participants and recruitment}
We recruited 50 participants in total (24 pairs and 2 individuals). Participants were required to complete a pre-screening eligibility survey that verified whether they met the inclusion criteria of the study prior to providing their informed consent and demographic information. The inclusion criteria for participation included: 1) Be 18 years or older; 2) Reside in the U.S. and be fluent in English, 3) Have at least two smarthome devices, 4) Can participate in the study on Zoom (optionally with a friend/family member who meets the above eligibility criteria 1-3). Our study was approved by our university Institutional Review Boards. We advertised through \edit{contacting participants from previous studies who expressed an interest in participating in future research, through} word-of-mouth, recruitment emails to relevant organizations and listservs, social media posts, and phone calls to people within our social networks, asking them to invite people they knew who were eligible. The recruitment process started in July 2021 and ended in April 2022. Table-\ref{tab:Demogrphic} shows the participant pairs' IDs, their gender information, relationships, \edit{and the smart devices they owned}. Overall, we recruited a sample of 50 participants (24 pairs of smarthome device owners and someone they knew, and two individuals). \edit{Our participants who self-reported their age fell into the following age ranges: 14\% were between 21-24 years, 58\% were between 25-34 years, 2\% were between 45-54 years, and 2\% were more than 55 years. The remaining participants did not report their age but were approximately in the 25-44 age range.} 
Most (56\%, N = 28) of our participant pairs were friends, whereas 28\% (N = 14) of them were \edit{life} partners, and 12\% (N = 6) identified themselves as siblings. \edit{Our participants reported (N = 36) owning a variety of smart devices: 42\% had smart TVs, 34\% had intelligent personal assistants (smart speakers), 26\% owned smart lights, 22\% had smoke detectors, 20\% owned smart power outlets or switches, 12\% had security cameras, 4\% reported owning motion sensors, and 4\% had water leak sensors. The remaining participants (N = 14) did not provide their smarthome device information.}


\renewcommand{\thetable}{\arabic{table}}

\begin{table*}[ht]
  \centering
  \footnotesize
\caption{Participant Profiles}
  \label{tab:Demogrphic}
\begin{tabular}{ c  c  c  c  p{11cm}} 
 \hline
 \textbf{Pair ID} & \textbf{Participant ID} & \textbf{Gender}  & \textbf{Relationships} & \textbf{Smart Devices Owned} \\ \hline \hline 
 G1 & P1, P2  & M, F & Siblings & --\\ 
 G2 & P3, P4  & M, F & Significant others  & --\\  
 G3 & P5, P6  & M, M & Friends  & --\\
 G4 &  P7, P8   & M, F  & Friends  & Light, speaker, outlet;  Light, speaker\\ 
 G5 &  P9, P10   & M, F & Significant others & Light, speaker;  Light, speaker\\ 
 G6 &  P11        & M &  --  & \\ 
 G7 &  P12, P13   & M, F  & Significant others  & --\\ 
 G8 &  P14, P15   & F, M  &  Significant others & --\\ 
 G9 &  P16        & F &  --  & \\ 
 G10 &  P17, P18   & M, F & Significant others  & --\\ 
 G11 &  P19, P20   & F, F & Friends & Speaker, television; Speaker, outlet \\ 
 G12 &  P21, P22   & F, M &  Friends & Speaker, outlet; Speaker, outlet\\ 
 G13 &  P23, P24   & M, M &  Friends & Light, doorlock; Light, doorlock\\ 
 G14 &  P25, P26   & M, F &  Significant others & Speaker, television; Speaker, security camera, smoke detector\\ 
 G15 &  P27, P28   & M, F &  Significant others & Light, speaker, security camera, doorbell; Light, speaker, security camera\\ 
 G16 &  P29, P30   & F, F &  Siblings & Thermostat, smoke detector, doorbell; Thermostat, smoke detector, doorbell\\ 
 G17 &  P31, P32   & M, F &  Friends & Smoke detector, television; Doorbell, television\\ 
 G18 &  P33, P34   & F, F & Friends & Smoke detector, television; Light, speaker, outlet, smoke detector\\ 
 G19 &  P35, P36   & F, F & Siblings  & Doorbell, television; Doorbell, television \\ 
 G20 &  P37, P38   & M, M & Friends & Light, speaker, outlet; Light, television\\
 G21 &  P39, P40   & M, M &  Friends & Thermostat, doorbell, speaker; Thermostat, doorbell, speaker\\
 G22 &  P41, P42   & F, M &  Friends & Light, television; Light, television \\ 
 G23 &  P43, P44   & F, F &  Friends & Light, thermostat, smoke detector, toy; Light, smoke detector\\ 
G24 &  P45, P46   & F, F &  Friends & Security camera, motion sensor, smoke detector; Light, smoke detector, television\\ 
G25 &  P47, P48   & F, M &  Friends &  Security camera, outlet, smoke detector, leak detector, television, toys; Outlet, smoke detector, television, watch \\ 
G26 &  P49, P50   & M, M & Friends & Speaker, television; Speaker, television\\ \hline \hline

\end{tabular}
\end{table*}

\subsection{Study \edit{Procedure}}
This study consisted of three distinct phases: 1) Semi-structured interviews about how and whether smarthome device owners currently share their devices with people who live outside of their homes, 2) Presentation of three hypothetical emergency situations in the context of smarthomes followed by semi-structured interview questions to reflect on whether sharing smart devices would be a good fit for those situations (or not), and 3) A think-aloud app interface co-design session with probing questions to understand how a smarthome co-monitoring app could help in emergencies similar to the hypothetical scenarios. We conducted 26 co-design interview sessions with 50 smarthome device owners (24 pairs who knew each other and two individuals) that took place remotely on Zoom. In each of the co-design sessions, participants were asked the same questions by researchers. Afterwards, we emailed each participant a \$20 Amazon gift card to thank them for their time. The \edit {entire} co-design interview session took from forty minutes to one and a half hours to complete, \edit{with an average duration of fifty minutes.} \edit {While the semi-structured interview part (the first two phases) took between fifteen to thirty minutes, the co-design part lasted from twenty-five minutes to an hour, with an average duration of 40 minutes}. Below, we discuss our study session design in more detail.

\subsubsection {Initial interviews}
In this initial phase of our co-design interview sessions, our goal was to understand participants' current approaches to managing their smarthome devices and their general motivation for sharing their smart devices with people outside their homes. We started this phase by asking them about the smarthome devices that they used in their homes. We also asked participants whether they currently shared their smart devices with anyone who lives outside of their homes, whom they shared with, and how they shared.


\subsubsection {Hypothetical scenarios}
We then presented storyboards to our participants that had images of three hypothetical smarthome emergency situations: 1) The first scenario \edit{(Appendix A)} depicted a smarthome device owner away from home who missed a notification from their smart smoke detector, leading to a fire; 2) The second scenario \edit{(Appendix B)} illustrated a situation where the smarthome device owner was on a camping trip with no cell signal. Their mom received a motion detection notification, checked the camera feed to confirm a potential break-in, and alerted the authorities; 3) The third \edit{scenario} 
 \edit{(Appendix C)}
presented a situation where a daughter had a friend over and they lit some candles. The smoke detector sent a notification to their mom's phone (as the device was shared with her), and the mom accessed the daughter's camera to confirm the fire incident and overheard their private conversation. \edit{The scenarios were drafted to describe two common property threats which can be monitored by IoT devices, and to introduce the idea of co-monitoring.} \edit{Furthermore, the inclusion of the third scenario was critical to surfacing and examining potential privacy threats of such a system.} We asked participants probing questions to learn about their reactions to these scenarios, as well as their suggestions as to what actions they would take to avoid these situations from happening. \edit{ Overall, the presentation of these three different scenarios not only motivated our participants to think about the potential benefits of allowing device sharing but also encouraged them to think about the privacy implications of it. Thus, participants could consider both convenience and privacy which helped them design the interfaces to support the benefits and alleviate the drawbacks. }

\subsubsection {Co-design session}
Finally, in phase 3, participants engaged in a co-design exercise where we asked them to brainstorm different solutions for their concerns in co-monitoring smarthomes for emergencies. We also asked participants to think through different emergencies that could happen in a smarthome and to design a mobile app interface to mitigate the issues. Throughout this part of the study, we encouraged the pairs to think aloud \cite{guan_validity_2006} as they designed the interface. We used Google slides for the design, as it is a low-tech technology that required little user training ~\cite{noauthor_benefits_2020} and could be easily shared between the researchers and participant pairs. Participants were given an empty mobile app structure with interface design elements (e.g., button, icon, text box, etc.). After they completed their app interface, we asked them to walk us through the interface design and discuss their rationale behind each feature. The study sessions were concluded by answering any final questions the participants may have had regarding our study. 

\setcounter{table}{0} \renewcommand{\thetable}{\arabic{table}}
\begin{table*}[ht] 
\centering
 \footnotesize
 \caption{Structure of Co-design Sessions with Sample Questions}
  \label{tab:interview-questions}
\begin{tabular}{ |p{3.5cm}|p{13.5cm}|  }
 \hline
\vspace{-1pt} \textbf{Structure} & \vspace{-1pt} \textbf{Sample Questions}  \\ 
\hline

\vspace{-1pt} \textbf{Phase-1: Semi-Structured \newline Interview} &
\begin{itemize}[leftmargin=0.4cm]
  \item  \textit{Please describe the smarthome devices that you use currently and how you use them at your home?}
  \item \textit{Do you share your devices with any people outside of your home? Why and when are you doing so?}
  \item  \textit{Do you share your smarthome devices with each other? If yes, how? }
 \end{itemize} \\ \hline

\vspace{-1pt} \textbf{Phase-2: Hypothetical \newline Emergency Scenarios \newline in Smarthomes} &
\vspace{-1pt} \textbf{ Scenario\#1:} A homeowner ignores her home smoke detector's notification and house fire is not prevented. 
\begin{itemize}[leftmargin=0.4cm]
  \item  \textit{Assume that you have a smart smoke detector, if you were the smarthome device owner in this scenario, and missed a notification from your smoke detector, how would this make you feel?}
  \item \textit{Is there anything you would think to do to try and avoid this situation?}
 \end{itemize} 

\vspace{-1pt} \textbf{Scenario\#2:} A mom watches a burglary-in-progress on daughter's home camera and stops it by calling 911.
\begin{itemize}[leftmargin=0.4cm]
  \item \textit{If you were the smart device owner in this scenario, how would you feel about sharing your smart devices in this way?} 
  \item \textit{What kinds of benefits would you envision if you shared smart devices for emergency monitoring in this way?}
  \item \textit{What concerns would you have if you enabled your smarthome to provide notifications and share your devices to other people in this way?}
 \end{itemize} 

\textbf{Scenario\#3:} A mom gets notified for smoke and accidentally listens in on daughter's conversation via camera. 
\begin{itemize}[leftmargin=0.4cm]
  \item \textit{If you were the smarthome device owner in this situation, how would you feel if something like this scenario occurred?} 
  \item \textit{Is there anything you think you could do to avoid this situation?}
 \end{itemize} 
 
 \\ \hline

\vspace{-1pt} \textbf{Phase-3: App Interface \newline Design Activities} &
\begin{itemize}[leftmargin=0.4cm]
\item \textit{Please list out the potential problems and their solutions as features that you might like in an app interface to configure sharing your smarthome devices to manage emergencies at your smarthome.}
\item \textit{Please design an interface that lets you share the smarthome devices with your emergency contacts, while addressing some of the concerns you raised. }
\item \textit{Please walk us through your interface design and explain how it would work when the two of you were sharing your smarthome devices with each other. }
\item  \textit{Can you tell me a story similar to the scenarios we showed previously to demonstrate what would happen in an emergency based on the controls you have in your designs?} 
 \end{itemize} 
\\ \hline
\end{tabular}
\end{table*}

\subsection{Data analysis approach}
The first author conducted the co-design interviews with the help of two research assistants. The co-design interviews were audio and video recorded and then transcribed. We also collected the Google slides as design artifacts participants produced. After completing the transcriptions, we conducted a grounded thematic analysis using Braun \& Clarke's \cite{braun2006using} six-phase framework to identify emergent themes. We first read through each transcript and visually analyzed the design-based artifacts to familiarize ourselves with the data. The first two authors discussed the transcribed content and designs to create the initial codes. During our initial coding, we also highlighted several important dimensions that seemed to be the most influential. We identified our emergent themes, always allowing the flexibility for new themes to emerge. While the first author coded all the transcripts, the second author re-coded some transcripts using the same set of codes but allowed the new codes to emerge. Every time a new code emerged in this step, the first two authors discussed until an agreement was reached on whether it is actually novel and worth adding to the codebook. Once they both agreed upon the newly emerged codes, the first author re-coded all the transcripts to include the new code. Therefore, our coding step was an iterative process where the first two authors frequently met to discuss the codes and form a consensus. 

After completing our coding, the first two authors worked with all co-authors to conceptually group the codes into cohesive themes that are aligned with our overarching research questions. For our RQ3, we not only coded the transcripts of the app interface design phase but also coded the design-based artifacts for each feature. The total count of the codes that appear in some themes can be greater than the total number of participants, as we double-coded the participant responses. For example, in our RQ1 codebook, some participants identified several benefits in co-monitoring their smarthomes in case of emergencies. Therefore, the count of the number of coded statements that co-monitoring could be beneficial exhibited a value that totaled more than 100\% of our participants.

\section{Results}
We now present the themes that emerged from our qualitative analysis. The participant’s quotations are identified with their Pair ID (G1, G2,...), Participant ID (P1, P2, ...), and their gender information.

\subsection{Potential motivations and benefits of sharing smarthome devices for in-home emergency co-monitoring (RQ1)}
\edit{While existing co-monitoring was not a selection criteria for this study, a number of participants reported this already occurring.} Thirteen (26\%) participants already shared their smarthome devices with trusted people who live outside of their homes or had access to others' devices. They mostly \edit{(N=10)} shared their home security devices with their family members who did not live with them, \edit{ but live in the same town,} to help them watch over their homes. For example: 

\begin{quote}
\edit{\textit{“At my parents' home, I have a Ring doorbell and Ring camera system that is shared with both my parents and myself, in case they're having a super busy day and someone breaks in. I have access to the cameras so I can see, and possibly call the authorities before my parents even noticed something happened.”} - G15, P27, Female}
\end{quote}

\noindent
\edit{A few other participants (N=3) shared their smart devices, such as smart door locks, with their extended family or trusted friends to enable easy access to their homes.}  

\begin{quote}
\edit{\textit{“So for our locks, we shared with my sister and my brother-in-law so that when they need to come in, they could use it... Sometimes the Ring alarm we also set it up, more family members or friends as also as notification wise, so they can be notified in case we're away and we don't have service."} - G7, P12, Male}
\end{quote}

\noindent
While the majority of our participants did not currently share, \edit{10\%, N=5 stated that they were not comfortable with co-monitoring at all}, while 64\%, N=32 indicated they might be willing to share devices for the purpose of emergency co-monitoring due to the potential benefits they could envision. These benefits encompassed practical expectations of having faster or better response with the help of other people, and a more general sense of safety and security that such co-monitoring would provide, which we describe in more detail below.

\subsubsection{Provides a sense of safety and security}  
Participants mostly envisioned that allowing their emergency contacts to co-monitor their smarthomes would give them a sense of safety and security in case of emergencies. 88\%, N=44 participants mentioned that having another person remotely keeping an eye on their home could save them from potential life threats and so it would \textbf{ensure their personal safety}. 

\begin{quote}
\textit{“I also agree \edit{[with P1]} like the fact that this household co-monitoring is for sure beneficial, like the implication of safety. I see this [co-monitoring] definitely would save lives.”} - G1, P2, Female
\end{quote}

\noindent
Participants imagined different scenarios when co-monitoring wou-ld be useful in case of emergencies. For example, when they were at work, but their kids or pets were home with babysitters or pet sitters, emergencies could still occur. Co-monitoring could lead to emergency contacts being able to immediately come to the house to help kids and pets from being injured or worse. Participants also often mentioned having multiple emergency contacts so that at least one person sees the notification and takes immediate action to save people's lives. Participants mentioned that they would rely on trusted family members and friends as their emergency contacts because these were people they felt already had a sense of responsibility in making sure they were safe.


\noindent Twelve participants (24\%) also mentioned that having their emergency contacts co-monitor their smarthomes could \textbf{prevent material loss}, saving their homes or belonging from being damaged. For example, P33 stated:

\begin{quote}
\textit{“The two things that come to my mind right away, is like saving my house from burning, saving from breaking in...it [co-monitoring] will save me from potential financial loss”} -G18, P33, Female
\end{quote}

\noindent Participants mentioned various emergencies that could damage a house, such as kitchen fires or water leaks, as well as loss of household items through theft. Significant cost savings would come from preventing such emergencies, or stopping them quickly before the damage is too great. 

Beyond the more concrete benefits of protecting themselves and their homes, participants also talked about \textbf{mental assurance} (18\%, N=9), the peace of mind that co-monitoring could provide. Participants often gave examples of situations where they might be on vacation, on a business trip, or were in a place with no phone signal. They could enjoy their trip or concentrate more on their work if someone trusted was watching over their homes. They frequently mentioned that it would allow them to not always have to be on high-alert, knowing that someone else would also be co-monitoring their homes when they were not available. 
\begin{quote}
\textit{“Another thing that comes to mind is peace of mind. You don't have to be always aware. Peace of mind part is when, like when you're busy, someone else can actually tackle a problem.”} -G20, P37, Male
\end{quote}

%

\subsubsection{Allows proxy monitoring when not home.}
\edit{In addition to the outcomes of co-monitoring}, participants discussed two ways in which having others co-monitor would lead to those benefits. First, in responding to the scenario where the device owner misses a notification, participants commented that co-monitoring would lead to a \textbf{quick response time} (44\%, N=22). They mentioned this could occur for numerous reasons -- if they were out of cell range, as in scenario 2, but even if they were home but sick or asleep. 


\begin{quote}
\textit{“It’s a more of the reaction time is faster, especially if you’re out of town, and somebody who’s close to the home, they can act faster. Somebody who monitors the home can respond to it faster.”} -G8, P15, Male
\end{quote}

\noindent 
A few participants (14\%, N=7) also pointed out that through co-monitoring the emergency contacts could \textbf{verify or confirm the emergency}, and thus take action when needed. Participants often brought up situations where they get emergency alerts from their smarthome devices, but later they found out that the alert was a false positive, similar to scenario 3. However, having multiple home devices shared helped them to verify the threats and decide whether they needed to act upon it or not. For example, \edit{P11 wanted to verify the seriousness of the threat by using a camera before taking any actions.} 



\begin{quote}
\edit{\textit{“So, if there was a screen where I can look at, let's say, a camera, So if there's a smoke detector going off, I want to check with the camera to make sure that there actually is a fire, and then some sort of button to alert emergency services"}} -G6, P11, Male
\end{quote}

\noindent 
Verification was particularly important if any automated actions could occur following a smart device alert, such as notifying authorities. Participants acknowledged that smarthome devices produce many false positives, and thus waste the time of emergency responders if there is a false alarm. For our participants, they saw their emergency contacts as an intermediate between an alert notification and a call to the authorities.


\subsection{Potential drawbacks and barriers of sharing smarthome devices for in-home emergency co-monitoring (RQ2)}
\edit{While many participants discussed potential benefits of co-monitor-ing, all} also raised concerns, some even seeing their concerns as being a barrier to adoption of such a system.

\subsubsection{Creates interpersonal issues}
Scenario 3 prompted participants to think about privacy, and indeed a common concern expressed by participants (64\%, N=32) was that having someone co-monitor their smarthome \textbf{might invade their personal privacy}. Participants often mentioned that they would feel uncomfortable if their partners or friends would come over while their parents or siblings were keeping an eye on them through indoor monitoring devices. Some participants also envisioned scenarios when they would talk over the phone and share their personal details, health information or even credit card information to someone, and the emergency contacts might overhear this sensitive information. Similarly, they also were concerned that indoor cameras could pick up their confidential emails, messages or bank login credentials which could be accidentally shared with their emergency contacts.  



\begin{quote}
\textit{“I think it violates some of the information privacy for me here. So, yeah, my password can be seen you know. Suppose I am sharing my living room cam with my brother. And he could just see my login info. So, this is like a privacy issue.”} - G2, P3, Male
\end{quote}

\noindent 
Beyond privacy, smarthome device owners (32\%, N=16) also expressed that asking emergency contacts to help them monitor their homes might \textbf{be a burden} for them. Participants often brought up situations similar to scenario 3, where the smoke detector alert was received because of a non-emergency situation. Some also imagined situations when motion detectors could get triggered by mailmen, pets, or even wild animals in the backyard. In such cases, the emergency contacts could be notified from the home camera or motion detection too often. Thus, co-monitoring would take effort, and could also be stressful and cause anxiety for their friends and family, particularly if there are too many false positives. 

\begin{quote}
\textit{“I think maybe worrying people unnecessarily, I think if it was, like just mentioned, false alarms. So you know, if there are notifications from the app that are not about emergencies that might be concerning to people that I know.”} -G22, P41, Female
\end{quote}

\noindent 
This theme contrasted with their own mental assurance from co-monitoring as they acknowledged that the mental burden of monitoring would be shifted to others.

\subsubsection{Creates security problems}
While most participants were concerned for potential privacy violations because of situations similar to the scenario 3, some also brainstormed additional scenarios where smarthome co-monitoring would cause more harm than benefits. One-third of the participants (32\%, N=16) suggested that sharing smart devices with outsiders \textbf{might cause unauthorized access to their homes}. In most cases, these scenarios involved user error. For instance, they envisioned a potential scenario where they might accidentally allow a wrong or untrustworthy person to have access to their smarthome devices. Some imagined other situations when this access misuse could take place, such as when the emergency contact forwards their device access URL or code to a third person. Finally, participants mentioned unauthorized access could also occur if the emergency contact loses their phone, and a third party could gain access to the home. 
\begin{quote}
\textit{“A concern that I would have is just if somebody else who, for example, wasn't my mom was able to have access to her phone and get that same notification. In that case, then it wouldn't be the person that I intended to share it with, which is my mom, would be whoever is able to have access to her phone, and may be able to see that information.”} -G12, P21, Female
\end{quote}

\noindent The idea was that if access could be shared, it might be shared accidentally or in the wrong way. In other words, these participants felt that the capability of giving access to their smarthome devices could create an unintentional ``back door'' into their physical homes.

Some participants (16\%, N=8) were also concerned about potential \textbf{cybersecurity attacks} that might occur when smarthome devices were shared with others. For instance, participants were concerned that their home security system could be hacked or their home network connection could get compromised as a result of co-monitoring their smarthome devices. 

\begin{quote}
\textit{“So sharing the devices would cause more risks, because if somehow our smarthome system is hacked from their phones, and our information gets leaked, then it might not be that useful to us.”} -G26, P50, Male
\end{quote}

\noindent To sum up, participants mostly were concerned about their personal privacy when it came to co-monitoring their smarthome devices with people outside their homes. We also observed that while participants \edit{could envision benefits of co-monitoring during in-home emergencies}, the drawbacks were mostly because of potential misuse of co-monitoring that might happen in non-emergency situations. \edit{Many participants mentioned} that the sense of physical safety, mental peace, and reduced risk of financial loss \edit{might} outweigh their concerns.

\begin{quote}
\textit{“It's like, there's so many pros and cons. The pros outweigh the cons as sense of life safety outweighs any nuisance part of it in this situation.} -G1, P2, Female
\end{quote}

\noindent However, for others, the privacy risks led to them to be unwilling to share their devices at all. 
\begin{quote}
\textit{“I feel like probably not, just because like privacy issues that might come up with that.} -G11, P20, Female
\end{quote}
\noindent Therefore, an important finding from our interviews was that an individuals privacy calculus \cite{princi_out_2020} of weighing the benefits (e.g., personal safety, prevention of material loss, and mental assurance) versus the potential privacy and security risks of co-monitoring led to their decision to adopt or not adopt such systems.


\subsection{Considerations for designing a collaborative smarthome emergency co-monitoring system (RQ3)}
In the co-design stage of the study, participants discussed a number of features they considered important for a co-monitoring system. These features focused on flexibly configuring \edit{device sharing to achieve co-monitoring}, while also mitigating their privacy concerns.

\begin{figure*} 
\begin{subfigure}[ht]{.23\linewidth}\centering
  \includegraphics[width=\textwidth]{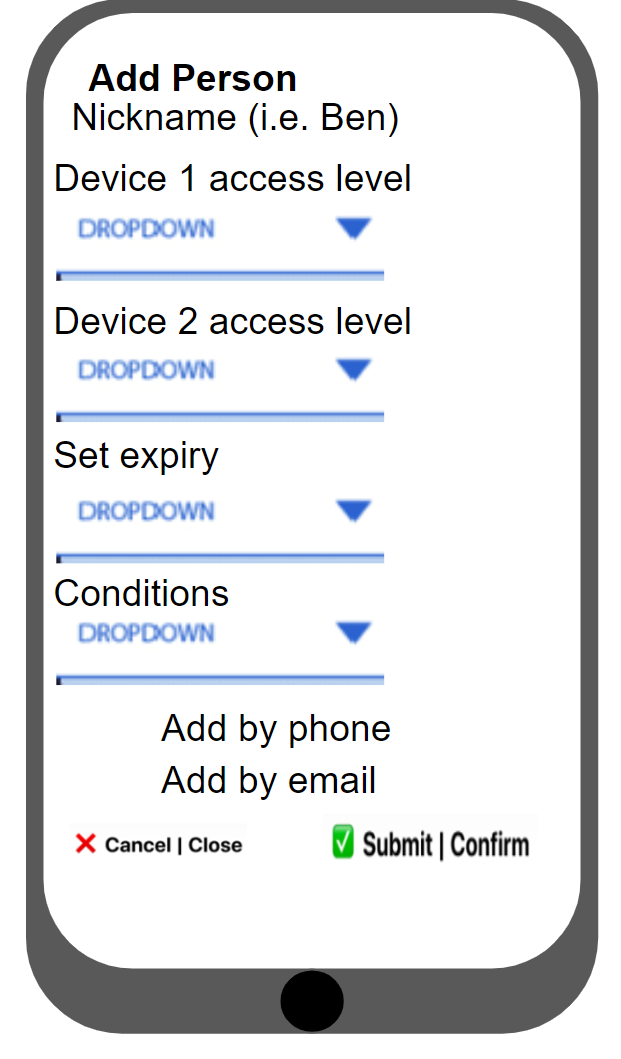}
  \caption{}
\end{subfigure}%
\begin{subfigure}[ht]{.23\linewidth}\centering
  \includegraphics[width=\columnwidth]{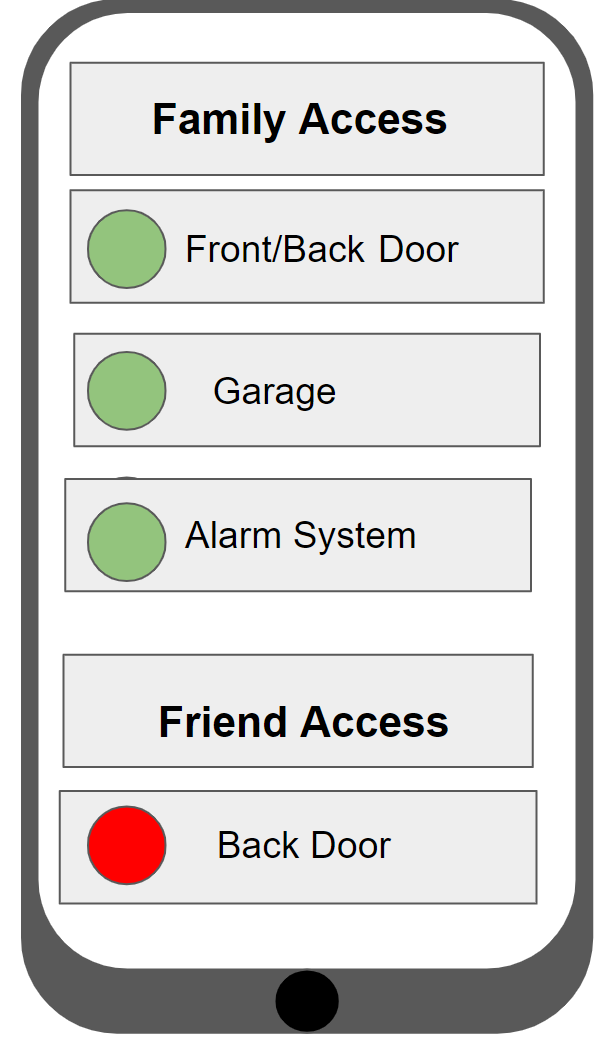}
  \caption{}
\end{subfigure}
\begin{subfigure}[ht]{.23\linewidth}\centering
  \includegraphics[width=\textwidth]{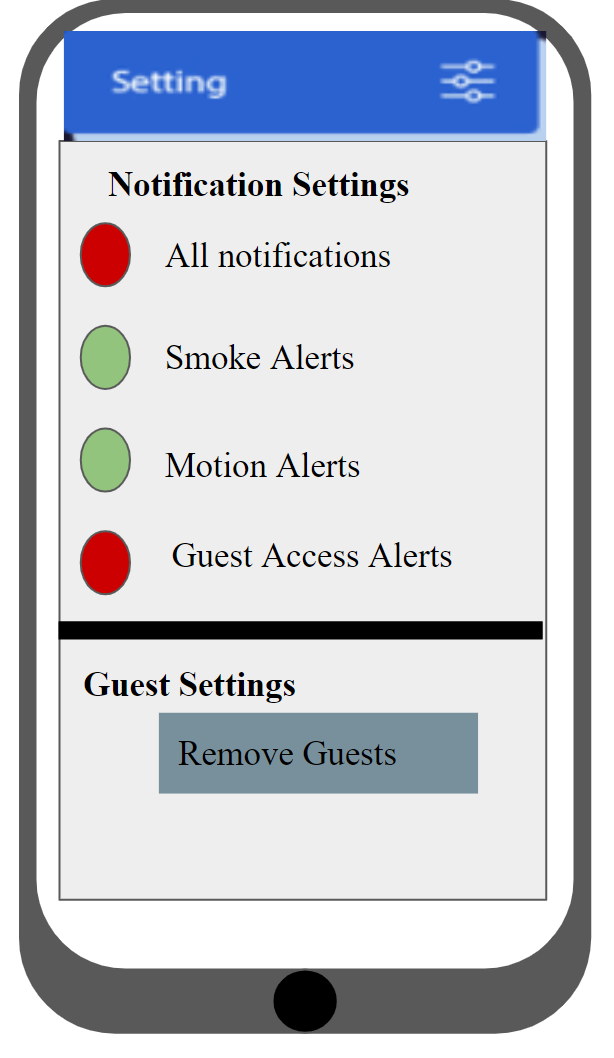}
  \caption{}
\end{subfigure}
\caption{\edit{Smarthome sharing features for emergencies: (a) Set a Time Access designed by Pair G21, (b) Specific Device Access for Specific People designed by Pair G1, and (c) Revoke Access feature designed by Pair G4}}~\label{fig:feature-1}
\end{figure*}

\subsubsection{Granular access controls}
All of our participants designed different kinds of granular controls for sharing the devices with various emergency contacts, based on different factors on which to base access. Approximately two-third of our participants (60\%, N=30) designed features that would allow them to \textbf{set a time schedule} for allowing their emergency contacts to have access to their smarthome devices. Participants often mentioned that they might be busy, sick, or sleeping during certain time periods where they would want co-monitoring to be active. They wanted to have that access revoked as soon as the assigned duration expired.  Figure-\ref{fig:feature-1}a depicts this feature, where pair G21 designed the ability to select devices and select the date and time when the access would expire.  
\begin{quote}
\textit{"This will take you to a different page where you can select specific time to access. You might want to give access only for specific time, as you know, you could be busy that time and you want someone to keep an eye.”} -G3, P5, Male
\end{quote}

\noindent 
More than half of our participants (56\%, N=28) also designed features that would allow them to share their devices with their emergency contacts only \textbf{when they are not home}. They also then wanted to have that access revoked automatically as soon as they are back home. Smarthome device owners designed this feature to avoid any potential privacy violation that might happen because of situations similar to our scenarios. 
\begin{quote}
\textit{“I think that it would definitely be more of a hey, I'm going on vacation, I won't be home type of thing. And I grant them access only for the allot of time that I'm not planning to be home. Up until I get home, they can see I arrived safely, and then they no longer have access to the cameras.”} -G15, P27, Male
\end{quote}

\noindent As many participants (40\%, N=20) mentioned that granting access for an unlimited period of time could cause major privacy and security issues, they also designed features to enable them to \textbf{easily revoke access} from specific emergency contacts at anytime. Figure-\ref{fig:feature-1}c illustrates a revoke access feature that the participant pair G4 designed. 
\begin{quote}
\textit{“I think that having some type of feature where you could force other users to log out of your system might be helpful. If you could have, like, a parent account and then your parents, like your mom, would sign on as a guest, you could revoke their access remotely”} -G4, P8, Female
\end{quote}

\noindent Around half (56\%, N=28) of our participants created features to \textbf{share only specific devices} with their emergency contacts (Figure-\ref{fig:feature-1}b). Participants also often mentioned that they might want to group certain devices together to ease configuration. For example, being able to easily provide access to both a smoke detector for fire detection and the indoor camera for visual verification. 



\noindent About half of the participants (52\%, N=26) also designed features that allowed them to share \textbf{only specific capabilities of devices} that may be useful in emergencies, rather than the full device. For example, participants mentioned they would want to share camera video but not audio, in order to allow for verifying emergency situations without potentially allowing eavesdropping on private conversations. 


 

\begin{figure*}
\begin{subfigure}[ht]{.23\linewidth}\centering
  \includegraphics[width=\textwidth]{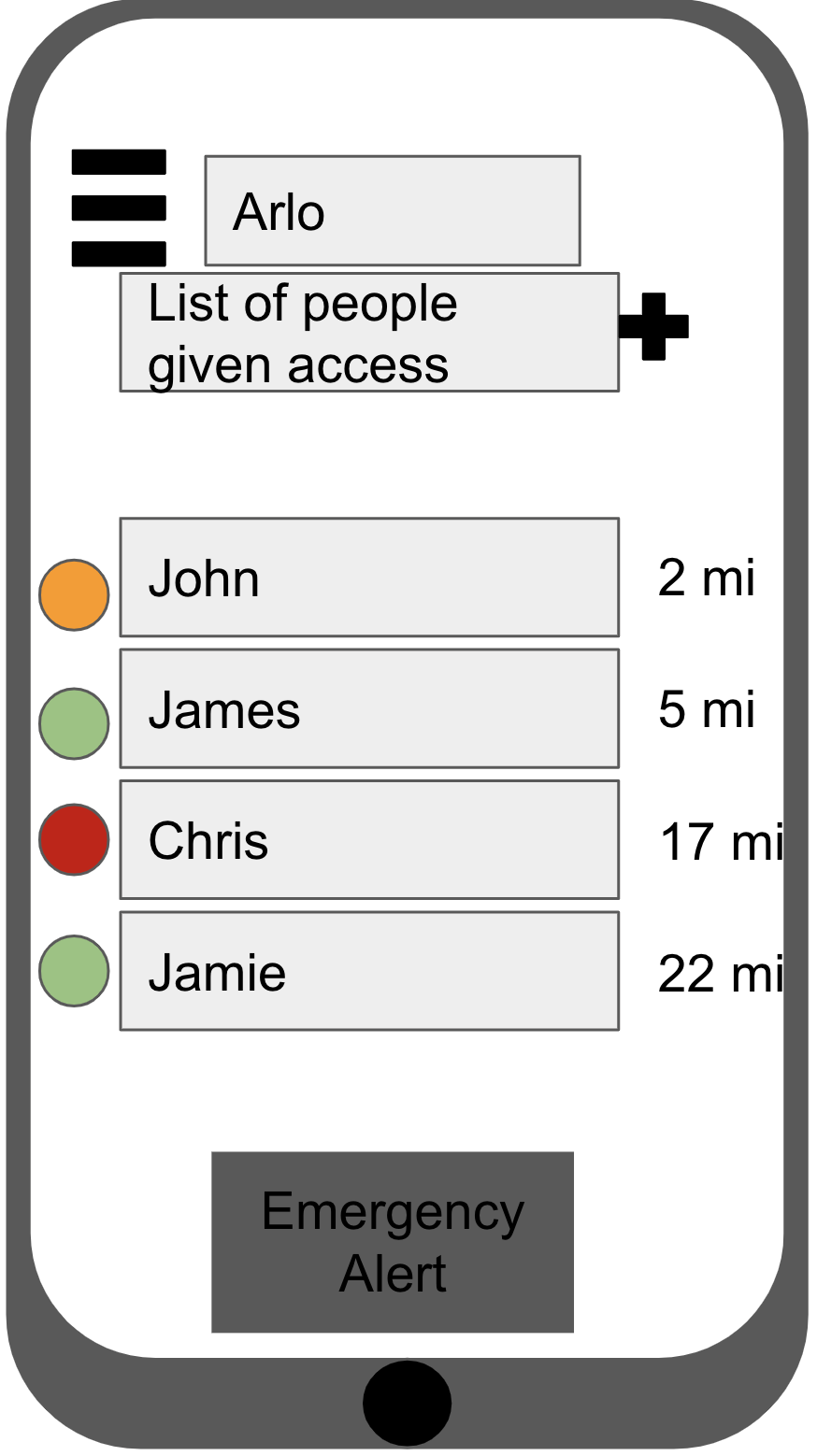}
  \caption{}
\end{subfigure}%
\begin{subfigure}[ht]{.23\linewidth}\centering
  \includegraphics[width=\textwidth]{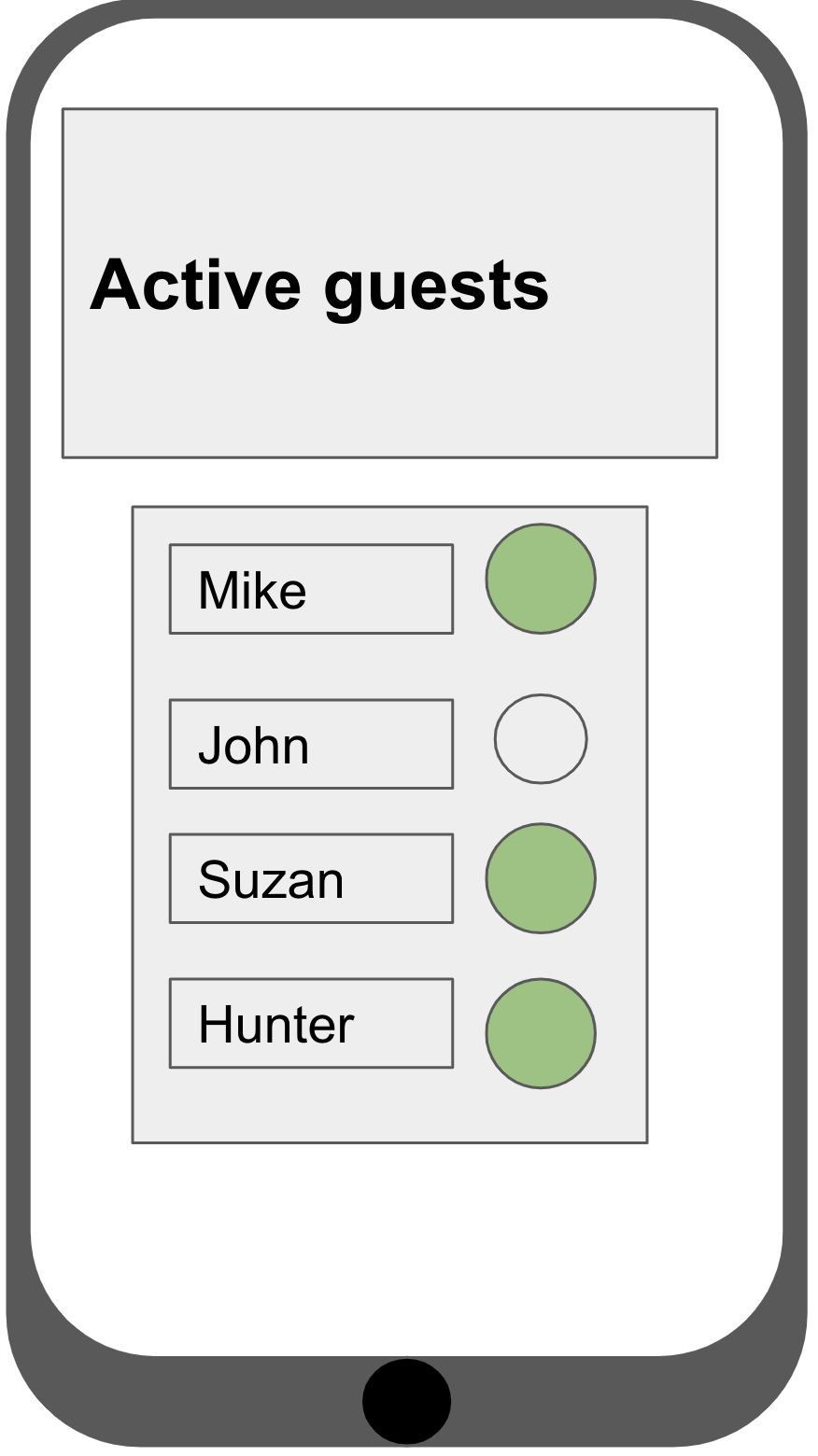}
  \caption{}
\end{subfigure}
\begin{subfigure}[ht]{.23\linewidth}\centering
  \includegraphics[width=\textwidth]{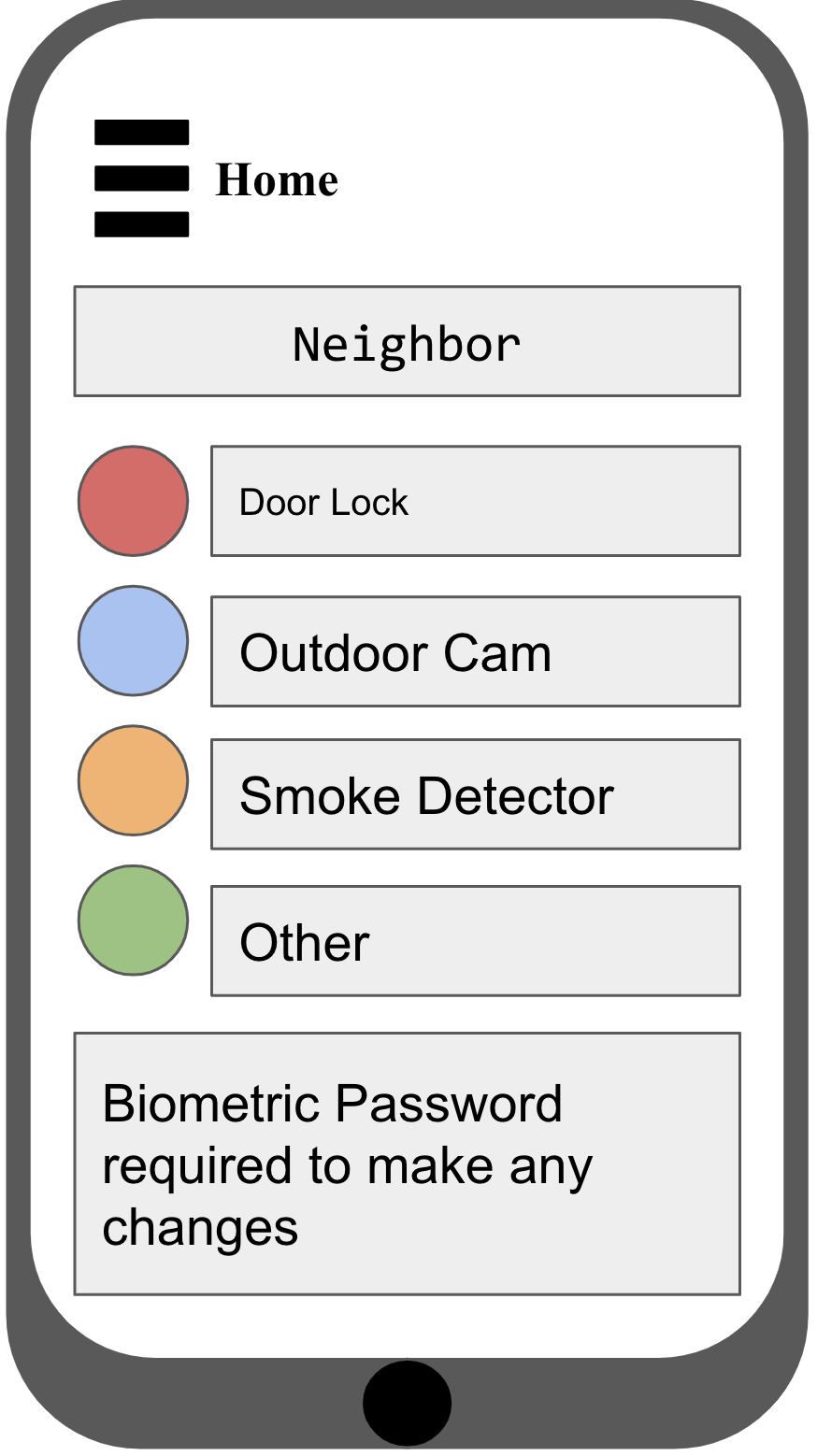}
  \caption{}
\end{subfigure}
\caption{\edit{Security features for smarthome sharing in emergencies: (a) Users' Locations designed by Pair G13, (b) Active Users Status designed by Pair G21, and (c) Change Confirmation designed by Pair G17}}~\label{fig:feature-2}
\end{figure*}

\subsubsection{Emergency contact status}
More than half of the participants also suggested some additional features to ensure transparency on the emergency contacts' statuses so that they can better organize and manage the co-monitoring in case of an emergency alert. For example, about a third of the participants (28\%, N=14) designed \textbf{location features} (Figure-\ref{fig:feature-2}a) to view the precise locations of all of their trusted emergency contacts with whom their smarthome devices are shared. Participants wanted to be able to check who is in the immediate vicinity of their home to respond to an emergency. They mentioned using this feature so that they could request an emergency contact to rush to their homes to deal with such an emergency first before they decide whether to call the emergency responders. 
 
\begin{quote}
\textit{“Potentially having like a view of where the people who are connected to your smarthome device are… in case something bad is happening or in case something goes wrong. And so just like a sharing location sort of feature...it could be helpful in the case that somebody is very close to your home. Maybe you're far away… your mom is nearby, and you can call her and say, like, please go there and see if we need to call emergency services. ”} -G12, P22, Male
\end{quote}

\noindent Some participants (24\%, N = 12) also designed a feature to check on \textbf{emergency contacts' activity} (Figure-\ref{fig:feature-2}b) to determine whether an emergency contact is \edit{ actively monitoring their smarthome devices through the co-monitoring app.} For instance, one group suggested using a colored dot next to a user's icon and name on the app might indicate their current status is active or not. Depending on this status, smarthome device owners would decide whether to replace an inactive emergency contact with another person from their trusted circle who can have time and willingness to actively co-monitor their homes for timely emergency assistance.

\begin{quote}
\textit{"We can have some feature that can show that who are actively using this one \edit{[co-monitoring app]} or who is not. So green glow spot for the active users or who are active at that point”} -G26, P50, Male
\end{quote}

\subsubsection{Safeguard against improper use}
As a number of participants had concerns over improper access of their devices, participants then designed features to help mitigate these concerns and allow them to monitor or protect against such risks. Thus, some (20\%, N = 10) introduced a feature to \textbf{confirm any changes made on emergency notification settings} (Figure-\ref{fig:feature-2}c), either by the smarthome device owner or the emergency contacts. Participants often mentioned that users might accidentally turn off a notification of a security device or share a device with a wrong person by mistake. So, there should be a feature that would ask users to verify these important changes, perhaps with added security (i.e. biometric password), to make sure that users actually intend to and are authorized to make the changes. 
\begin{quote}
\textit{“I think it would also be helpful to have like a confirmation system, so you don't accidentally turn on somebody's notification, or like someone's access.”} -G4, P8, Female
\end{quote}

\noindent Participants (16\%, N=8) also wanted to have another security feature that \textbf{logs all activities of the smarthome devices}. For instance, the app should log the timestamps of all the actions made with the smarthome devices and also keep a record of the people who accessed those devices. The smarthome device owners can later check the device activity log that lists out all the actions taken by the emergency contacts to make sure the smarthome devices are not being misused.

\begin{quote}
\textit{“Last app screen would just be log of events that are happening with a device. So like, not only does your device and notification, if it detects motion in the front door, it also has the log, you can check. That says that you can see a timeline of, oh at this time someone went into the house”} -G13, P24, Male
\end{quote}

\subsubsection{Ensure the emergency notifications are noticeable and received}
Most of our participants also suggested additional features that would ensure that the emergency notifications are difficult to ignore or overlook. They had various suggestions for how to make those notifications stand out from all the others on their phones.

\begin{quote}      
\textit{“The emergency alerts gonna have a different notification than per se, like, movement on the camera. So there'll be a sense of urgency.”} -G1, P1, Male
\end{quote}

\noindent Many participants (44\%, N=22) recommended a feature for \textbf{unique and loud alert sounds} based on the emergency type. In other words, they wanted different sounds for different types of emergency, such as fire or break-in, so that the device owner and the emergency contact can distinguish those notifications from others. Participants (24\%, N = 12) also suggested that the co-monitoring app should send notifications in such a way that it can \textbf{override the phone's Do Not Disturb or silent mode}.

\begin{quote}
\textit{“I would  make it an app that can bypass Do Not Disturb that way even if my phone was on silent, on Do Not Disturb.”} -G15, P27, Male
\end{quote}

\noindent A few (12\%, N=6) suggested sending \textbf{persistent notifications} during emergencies, such as every 30 seconds, until the notifications are acknowledged and acted upon. 

\begin{quote}
\textit{“That means my smoke alarm is going off frequently. If we don't open the app, we don't tend to the notification. It'll send us an alert every 30 seconds or every minute until we tend to it”} -G15, P28, Female
\end{quote}

\noindent Finally, some (20\%, N=10) wanted to switch how notifications were delivered, such as through a \textbf{direct phone call or text} rather than an app. Participants envisioned a situation when smarthome device owners or the emergency contacts are busy at work and do not have time to read app notifications. They thought phone calls and texts would lead to a better and faster response.

\begin{quote}
\textit{“I would probably rather it be like a call or like something that just from the notification itself. ...Some automated call.”} -G4, P7, Male
\end{quote}

\subsubsection{Automated emergency response}
\noindent Even though we asked our participants to design an app for in-home emergency co-monitoring, some participants (28\%, N=14) wanted features that would \textbf{directly send notifications or call emergency response} so that the situation could be handled in a timely manner by professionals.

\begin{quote}
\textit{“If you are away from home, it's definitely an emergency situation. We definitely want this to be notified to the 911 or emergency services so that they can act in time”} -G8, P15, Male
\end{quote}

\noindent Other participants (20\%, N=10) suggested that the co-monitoring app could send notifications \textbf{through an intelligent emergency detection system}. In other words, they wanted the app to have built-in intelligence to decide whether the situation is actually an emergency or not before sending notifications or notifying emergency response directly. 
\begin{quote}
\textit{“I think if there’s a device that would send kind of smoke or gases in the atmosphere analyze that, you know, based on the severity then it will send like an alarm or notification to the police department”} -G3, P5, Male
\end{quote}



\noindent To sum up, our participants designed features to help smarthome owners and emergency contacts co-monitor their homes together to ensure safety and security in emergencies in an attempt to mitigate the  downsides that come with co-monitoring. Participants did not just design features to resolve the issues they saw in the hypothetical scenarios; they also came up with additional scenarios and their corresponding solution features.

\section{Discussion}
In this section, we describe the implications of our findings in relation to prior work and provide design implications for designing smarthome emergency alert systems for co-monitoring purposes.

\subsection{Smarthome device owners are motivated to protect their homes}
The overarching result of our study is that people are motivated to protect their homes and do desire help in monitoring them. A number of participants requested what they would really like, which is a system that could automatically determine when there is a real emergency and alert emergency responders. However, automated emergency response systems or monitored alarm systems are more expensive to install than most smart devices, and still suffer from false alarms, resulting in collateral costs and wasted time for responders. 


A key motivation to adopt smart devices is to monitor the safety of one's home. And thus, despite this desire for automating emergency response, many participants recognized the benefits that co-monitoring could provide in sharing the responsibility of monitoring with trusted contacts \cite{badillo2020towards}.  \edit{Our study highlights important considerations for this use case, and the need to consider smarthome usage from a community perspective.} Co-monitoring may even be preferred over professionally monitored alarm systems, as those a device owner trusts could be provided access to other devices, such as cameras, to verify an emergency situation before taking action. If co-monitoring can provide sufficient context and accurate information, this will aid decision-making and reduce damage to life or property  ~\cite{engelbrecht_digital_2011}. Future research would be needed once such systems are deployed to determine how false positives occur and design to further mitigate them.

Our findings reinforce and expand upon those from prior work by Tabassum et al. ~\cite{tabassum_smart_2020} that has shown that people are already sharing their smarthome devices with people who do not live with them to help in taking care of the home. Thus, while all participants voiced privacy concerns regarding sharing their devices, for many, the potential to improve emergency response and the peace of mind that provided outweighed those concerns. \edit{By understanding the strong motivations of users to have help in emergency monitoring, smarthome system designers can provide features that allow users to weigh the adoption, use, and sharing of their devices against privacy intrusions that will arise when sharing devices with others.}

\subsection{Privacy is a major barrier for smarthome co-monitoring}
Our study confirmed that privacy concerns weigh heavy on the minds of smarthome device owners, and may impede adoption of co-monitoring applications. This is consistent with studies that investigated the sharing of smart devices with other people and identified privacy and security challenges ~\cite{brush_its_2012,jang_enabling_2017} as factors that influence the decision of whether or not to share a device. Similar privacy concerns have been found in prior work on device sharing ~\cite{zeng_end_2017,tabassum_smart_2020}, where participants were only willing to share devices with those they most trusted or when privacy intrusions were less likely, such as when they were not at home. Our participants expressed similar thoughts. Thus, users are likely to be more willing for their most trusted friends and family to be emergency contacts for co-monitoring, and less likely to share device access with other useful community members, such as nearby neighbors who may have easier physical access to one's home. \edit{These results further demonstrated that addressing privacy concerns, such as access by less trusted community members, may allow users to expand their use and sharing of smarthome devices in ways that provide desired outcomes.}

However, some of the solutions that participants designed to mitigate their concerns, or improve the use of co-monitoring, may introduce their own privacy issues. For example, a co-monitoring system that shares devices when one is not home would need some way to determine that, either through location or other sensing. This then requires additional data collection or inference. Some participants also suggested a way to check which of their contacts is available or nearby in the event of an emergency. Again, this requires data collection from the contacts themselves, which may impede their desires to be involved in such a system. In addition to the system itself now collecting location, the potential to share that location with other people, even if only in rare situations, can have unintended consequences, allowing friends and family to determine that someone was not where expected ~\cite{barkhuus_location-based_2003}. However, location-tracking services have a chance of success in such a system if users are given a simple option to turn off location tracking when it is not needed ~\cite{barkhuus_location-based_2003}. There is a delicate balance between enabling goals such as “safety” and “convenience” without crossing the boundary of making smarthome device owners feel a system is “privacy intrusive”. This balance is difficult to meet, especially given the fact that smarthome device owners may share their devices with multiple people, each with different values and roles (i.e, family, or friends).

One novel observation in our study is the extent that participants worried about privacy invasions from attackers or unauthorized third parties as a result of sharing for co-monitoring. Security attacks and unauthorized access of recorded information by third parties have been reported as a concern of smarthome devices in general ~\cite{denning_computer_2013,bugeja_privacy_2016}. Yet, our participants seemed concerned that an added system for co-monitoring, which does not actually increase or change data collection, could add to those concerns and increase security and privacy risks. Early adopters of such systems may need reassurance that their smarthome device and collected information is secure.  \edit{While none of our participants raised more serious privacy concerns, such as the potential harm of implementing smarthome co-monitoring in close relationships that were indicative of abusive power dynamics, we would be remiss in not considering this possibility ourselves. Indeed, several HCI researchers (e.g., ~\cite{leitao_anticipating_2019, parkin_usability_2020, rodriguez-rodriguez_towards_2020, slupska_threat_2021, mcdonald_privacy_2022, noauthor_ipv_2022}) have already highlighted the potential harms from smarthome-based surveillance technologies and the need for trauma-informed practices when developing technologies that may further exacerbate existing harm ~\cite{chen_trauma_2022}. Therefore, we include some of these recommendations in our implications for design.}

\subsection{Guidelines for designing a smarthome emergency co-monitoring system}
Our study provides insight into the features and mechanisms that would be needed in an app for sharing smarthome devices with people outside the home for emergency co-monitoring. Our findings suggest features that designers should consider in supporting smarthome device owners' needs, as well as mitigating their \edit{privacy and security} concerns.

\textit{Granular access control} 
By far the most strongly suggested feature was the need for \edit{granularly} managing access control, configuring who was an emergency contact and exactly when and what notifications and devices they could access. Participants designed features to customize the access levels based on their preferred schedule, device types and features, relationships with contacts and even their own location. Many of these features have been suggested by prior work in sharing smarthome devices more generally, both within and outside of the home ~\cite{tabassum_smart_2020}. For example, Tabassum et al, advocated for more time- and/or event-based access controls to support remote monitoring and notification requirements ~\cite{tabassum_smart_2020}. \edit{Our participants anticipated new ways of sharing and revoking smarthome access control that empowered smarthome device owners to effectively manage with whom, when, how long, and under what circumstances these trusted individuals would have physical and/or virtual access to their homes. In a sense, our participants anticipated and designed for four of the six dimensions (i.e., safety, trust, peer support, and collaboration) of Chen et al.'s \cite{chen_trauma_2022} framework of trauma-informed computing in their designs. Safety was considered when conceptualizing with whom and how to share smarthome devices for the purpose of protecting one's home. Trust was central to forming an emergency contact relationship, where participants acknowledged both the benefits and the burden of asking a trusted person outside of their homes to engage in co-monitoring. Therefore, features for managing access control have to be designed carefully to promote a collaborative sharing process, while balancing the desire to protect the smarthome device owners' privacy and safety.}

\edit{Yet, our} results also demonstrate how difficult it might be to simply use existing device-specific sharing interfaces to satisfy access needs, where different capabilities of multiple devices might be shared upon a particular kind of alert or notification from one device, at specific times. Configuring each device separately, outside the context of emergency response, would be difficult and error-prone for users. Smart device integration platforms that provide integrated apps for controlling multiple devices could provide co-monitoring configuration interfaces. However, providing a flexible access control system also makes such an interface more complex. And this complexity may be too burdensome or overwhelming for users, and instead go unused. Indeed, Zeng and Roesner, ~\cite{zeng_understanding_2019} built and deployed a prototype access control system with multiple options for sharing smart devices within a home, and found that much of that system went unused due to the complexity caused by both the granularity of the settings, and the number of different devices managed by the home. Designers will need to find a balance between enough flexibility ~\cite{alghamdi_micaa_2022} to meet user needs and sufficient simplicity and usability, to provide controls that easily support common use cases.

\textit{Transparency of sharing} Another challenge with current smart-home device sharing interfaces is their lack of transparency. Past studies have found that owners have difficulty determining what exactly is shared when they allow access to other people \cite{ur_current_2013}. Without that knowledge, users have to assume they are sharing everything, and as a result share only with those closest to them. Others resort to simply sharing full account credentials \cite{kim_challenges_2010}. Configuring for emergency response adds complexity with multiple factors on which access could be based, as users will want to balance co-monitoring needs with privacy. Yet, hopefully emergency alerts, and the resulting access, would happen infrequently. Providing transparency and a full understanding as to what can be viewed and accessed by others in such infrequent situations could be challenging. Thus, methods are needed to help both smarthome device owners, as well as emergency contacts, build up that understanding of how access will work. \edit{Again, applying a trauma-informed lens, transparency, in terms of logging and alerting smarthome device owners exactly when and how smarthome device control was enacted by one's emergency contacts, was how our participants further mitigated privacy, security, and safety concerns of such a system. Therefore,} beyond access control, our results suggest the following set of design guidelines specific to smarthome emergency co-monitoring.

\begin{itemize}
\item \edit{Improve user-awareness by enabling them to determine what emergency contacts are able to access and how they will interact as a co-monitor, to inform users' mental models of data flows in the system.}
\item \edit{Provide confirmation dialogs, so users can review and verify changes in access configuration.}
\item \edit{Provide indicators of activity, such as through logs, so users can determine what and when emergency contacts are accessing devices.}
\item Make emergency notifications stand out, such as through unique sounds or alternate delivery.
\item Make emergency notifications hard to overlook and ignore, such as through repeated delivery or overriding DND/silent modes.
\item Provide methods or guidance to emergency contacts on how to alert authorities when needed to ensure fast response.   
\item Provide methods for device owners to reach out directly to the best contact for a situation, such as through determining who is in the immediate vicinity of the home.
\item \edit{Provide a quick method for users to revoke access to devices.}
\item \edit{Provide additional safeguards such as creating systems/devices with privacy in mind to protect vulnerable users and also to relieve users of some of the responsibility of safeguarding their own privacy.}

\end{itemize}

\edit{For the last point, our participants did not explicitly consider the sixth dimension of trauma-informed computing in their designs, specifically that of \textit{intersectionality} \cite{chen_trauma_2022}, which acknowledges that some people experience forms of oppression that are not generalizable to the general population. Additional safeguards for vulnerable users might include embedding the contact information for the domestic abuse hotline in the help guidance of an app, as to raise victim's awareness that digital surveillance is a form of abusive control \cite{chen_trauma_2022}. Another safeguard might include a ``privacy'' or ``SOS'' mode, where smarthome device owners can turn off monitoring (without notifying the untrusted emergency contact) or reach out directly for help from the proper authorities. While such safeguards may not be relevant to typical smarthome device users, they could be critical to the safety of those in abusive situations; thus, worth implementing.}

\subsection{Limitations and future work}
This work used three scenarios regarding potential motivations for sharing access to smarthome devices in emergencies. These scenarios may have influenced participants' opinions of how smarthome device owners would or should share their smarthome devices and with whom in an emergency. These scenarios were hypothetical and may not reflect all potential real-world situations. \edit{Additionally, all participants participated in the co-design sessions, even if they were skeptical they would want to adopt such a system in the future.} Thus, there might be some differences between our participants’ responses and their real-life perceptions about sharing smarthome devices in different contexts. Therefore, future work should conceptualize new and different scenarios to minimize the impact of these scenarios on participants’ perceptions of smarthome co-monitoring. We also suggest using design fictions \cite{yao_ubiquitous_2019} as a way to envision new scenarios and use cases in which smarthome co-monitoring may be beneficial or problematic. 

As a \edit{small-scale} qualitative study, this work sheds light on various recurring themes that we identified in our co-design interviews and should not be seen as generalizable to all smarthome device owners in a statistically significant sense. We leveraged a broad, non-probability sampling frame; therefore, our participants may not represent the overall demographics of all smarthome device owners. As we did not collect detailed demographic information, e.g., race, socioeconomic status, education, and employment, we cannot characterize the diversity of our participants and future work should explore a broader set of users along a number of demographic dimensions \cite{shruti_privacy_2022, warford_sok_2022}. Likely due to our broad inclusion criteria, recruitment strategy, and participants' self-selection in participating in the study with someone they knew and trusted, we consequently did not engage, at least to our knowledge, with more vulnerable users, such as children, LGBTQ+ community, or victims of intimate partner violence. Future research should focus on these unique contexts and users to ensure their safety when developing smarthome co-monitoring solutions that could potentially facilitate unanticipated harms when trying to improve in-home security.

\section{Conclusion}
\edit{Our study explored shared emergency monitoring, a use case that is already driving smarthome device owners to share their devices with people outside of their homes. } We also investigated smarthome device owners' perceptions about the potential benefits, drawbacks, and design considerations for developing a smarthome system to facilitate co-monitoring with trusted emergency contacts who live outside of one’s home. Generally, participants felt that such a system would help ensure their personal safety, protect against material loss, and give them peace of mind by ensuring rapid response to emergencies and verifying potential threats. \edit{Yet, the study also demonstrated the complex privacy challenges that arise in such a collaborative system. Participants sought to address these challenges with granular access controls, in ways that current systems do not yet support.} Given the continued proliferation and adoption of various kinds of home safety and security smart devices, we believe co-monitoring is an important use case for smarthome systems. We will continue to build upon this work to examine how we can provide support for communities of users to \edit{safely and privately} help each other in monitoring and protecting their homes.

\begin{acks}
We would like to thank the individuals who participated in our study. We also thank Elizabeth Chan, Tailin Postema, and Supriya Nittoor for their assistance with conducting the study. This research was supported by the U.S. National Science Foundation under grants CNS-1814068, CNS-1814110, CNS-1814439, and CNS-1757884. Any opinion, findings, and conclusions or recommendations expressed in this material are those of the authors and do not necessarily reflect the views of the U.S. National Science Foundation.
\end{acks}





\bibliographystyle{ACM-Reference-Format}
\bibliography{sample-base}

\newpage
 \onecolumn
\appendix

\label{Appendices}

\begin{appendices}
\section{Hypothetical Scenario 1}
\begin{figure*}[h]
  \centering
  \includegraphics[width=\linewidth]{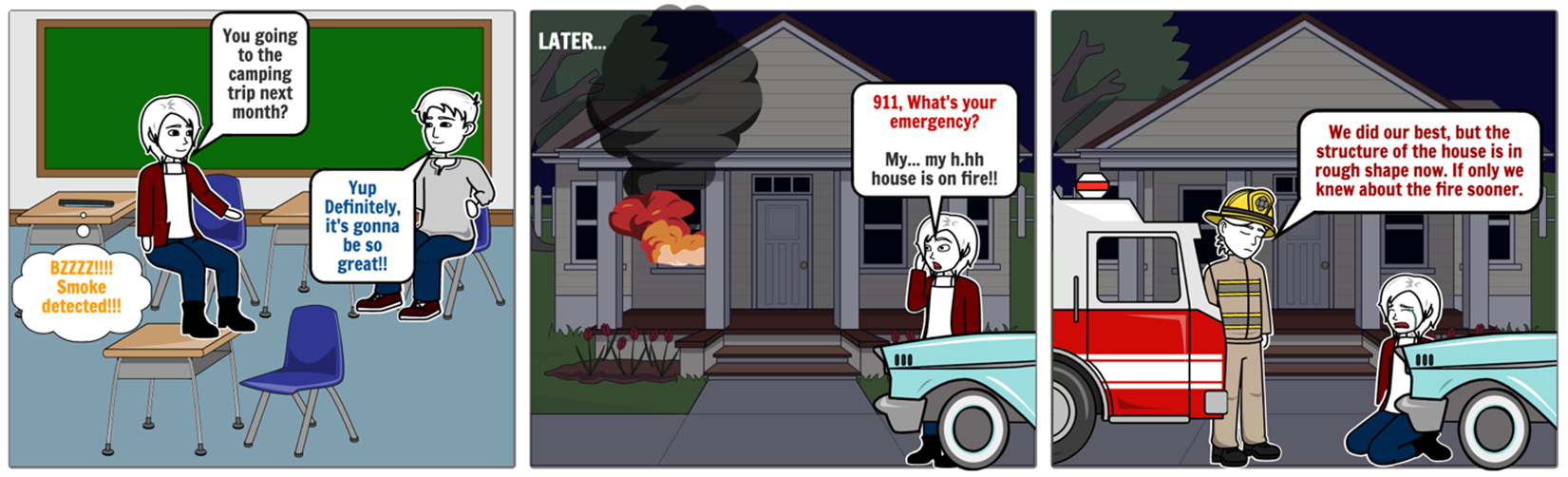}
\end{figure*}

\section{Hypothetical Scenario 2}
\begin{figure*}[h]
  \centering
  \includegraphics[width=\linewidth]{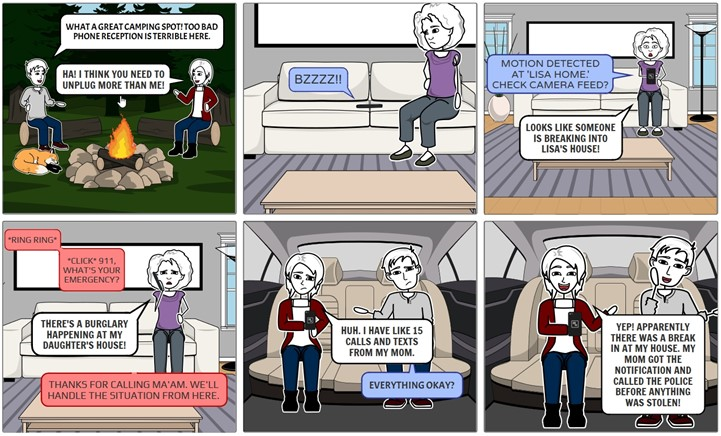}
\end{figure*}

\newpage
\section{Hypothetical Scenario 3}
\begin{figure*}[h]
  \centering
  \includegraphics[width=\linewidth]{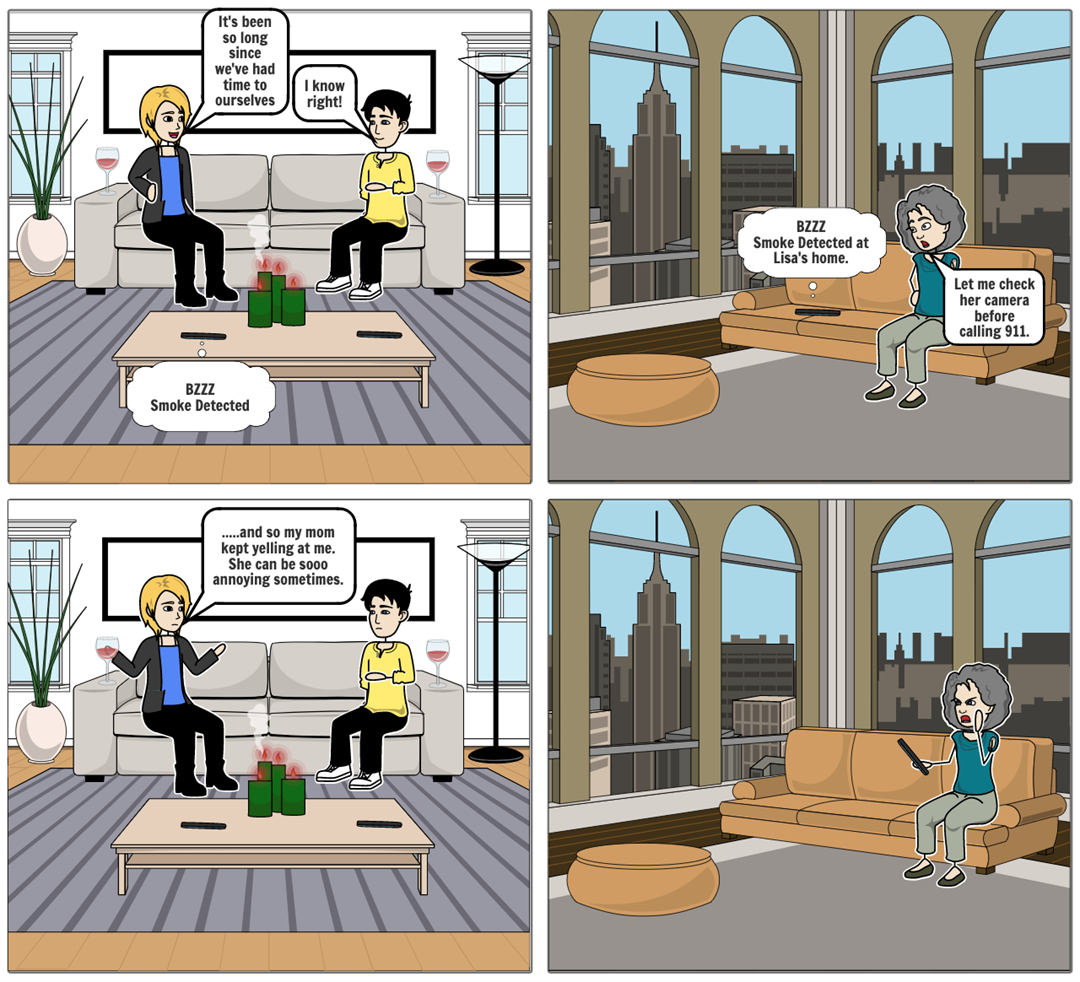}
\end{figure*}

\end{appendices}
\end{document}